\documentclass{nature_withfigs}

\usepackage{amsmath,amsfonts,amssymb}
\usepackage{wrapfig}
\usepackage{graphicx}

\title{Quantum gas of rovibronic ground-state molecules in an optical lattice}

\author{Johann G. Danzl$^1{^\ast}$, Manfred J. Mark$^1$, Elmar
Haller$^1$, Mattias Gustavsson$^1$, Russell Hart$^1$, Jesus Aldegunde$^{2}$, Jeremy M. Hutson$^2$ \& Hanns-Christoph N\"agerl$^1{^\ast}$}

\begin{document}

\newcommand{\WW}{\ensuremath{\mathcal{W}}}
\newcommand{\QQ}{\ensuremath{\mathcal{Q}}}

\newcommand{\new}[1]{\textit{#1}}

\newcommand{\ketbra}[1]{\ensuremath{| #1 \rangle \langle #1 |}}
\newcommand{\ket}[1]{\ensuremath{|#1\rangle}}
\newcommand{\bra}[1]{\ensuremath{\langle#1|}}
\newcommand{\braket}[2]{\ensuremath{\langle #1|#2\rangle}}

\maketitle

\begin{affiliations}
\item Institut f{\"u}r Experimentalphysik und Zentrum f\"{u}r Quantenphysik,
Universit{\"a}t Innsbruck, Technikerstra{\ss}e 25, A--6020 Innsbruck, Austria
\item Department of Chemistry, University of Durham, South Road, Durham, DH1 3LE, England
\end{affiliations}

\begin{abstract}
Control over all internal and external degrees of freedom of molecules at the level of single quantum states will enable a series of fundamental studies in physics and chemistry\cite{Carr2009,Krems2008}. In particular, samples of ground-state molecules at ultralow temperatures and high number densities will allow novel quantum-gas studies\cite{Goral2002qpo} and future applications in quantum information science\cite{DeMille2002}. However, high phase-space densities for molecular samples are not readily attainable as efficient cooling techniques such as laser cooling are lacking. Here we produce an ultracold and dense sample of molecules in a single hyperfine level of the rovibronic ground state with each molecule individually trapped in the motional ground state of an optical lattice well. Starting from a zero-temperature atomic Mott-insulator state\cite{Bloch2008} with optimized double-site occupancy\cite{Volz2006}, weakly-bound dimer molecules are efficiently associated on a Feshbach resonance\cite{Chin2008} and subsequently transferred to the rovibronic ground state by a stimulated four-photon process with $>$50\% efficiency. The molecules are trapped in the lattice and have a lifetime of 8 s. Our results present a crucial step towards Bose-Einstein condensation of ground-state molecules and, when suitably generalized to polar heteronuclear molecules, the realization of dipolar quantum-gas phases in optical lattices\cite{Baranov2008,Lahaye2009,Pupillo2008cmp}.
\end{abstract}

Recent years have seen spectacular advances in the field of atomic quantum gases. Ultracold atomic samples have been loaded into optical lattice potentials, allowing the realization of strongly-correlated many-body systems and enabling the direct observation of quantum phase transitions with full control over the entire parameter space\cite{Bloch2008}. Molecules with their increased complexity are expected to play a crucial role in future generation quantum gas studies. For example, the long-range dipole-dipole force between polar molecules gives rise to nearest-neighbour and next-nearest-neighbour interaction terms in the extended Bose-Hubbard Hamiltonian and should thus lead to novel many-body states in optical lattices in the form of striped, checkerboard, and supersolid phases\cite{Baranov2008,Lahaye2009,Pupillo2008cmp}.

An important prerequisite for all proposed molecular quantum gas experiments is the capability to fully control all internal and external quantum degrees of freedom of the molecules. For radiative and collisional stability, the molecules need to be prepared in their rovibronic ground state, i.e. the lowest vibrational and rotational level of the lowest electronic state, and preferably in its energetically lowest hyperfine sublevel. As a starting point for the realization of novel quantum phases, the molecular ensemble should be in the ground state of the many-body system. Such state control is only possible at ultralow temperatures and sufficiently high particle densities. While versatile non-optical cooling and slowing techniques have recently been developed for molecular ensembles\cite{Friedrich2009}, the achievable molecular phase-space densities are still far away from the point of quantum degeneracy. Here, we exploit the fact that high phase-space densities can readily be achieved for atoms and that atoms can efficiently be associated on Feshbach resonances to form molecules\cite{Chin2008} with minimal loss of phase-space density when an optical lattice is present. Subsequent state transfer to a specific hyperfine sublevel of the rovibronic ground state by means of a stimulated multi-photon process then preserves phase-space density and hence the quantum-gas character of the molecular ensemble. This approach is expected to allow the preparation of a molecular ground-state BEC\cite{Jaksch2002}.

A crucial ingredient for our experiments is the presence of an optical lattice. It provides full control over the motional wave function and prevents collisional loss. It allows us in particular to maximize the efficiency for initial molecule production and the efficiency for ground-state transfer. In addition, our preparation procedure directly leads to a lattice-trapped molecular many-body state, which forms the basis for the many-body and quantum information experiments envisioned\cite{Baranov2008,Lahaye2009,Pupillo2008cmp}. In the quantum gas regime without the use of an optical lattice, molecular state transfer to deeply-bound rovibrational levels of the singlet $^1\Sigma$ ground-state potential has recently been implemented for Cs$_2$\cite{Danzl2008Science} and KRb\cite{Ni2008}. For KRb, the rovibronic ground state was reached, resulting in a near-quantum-degenerate gas of fermionic ground-state molecules\cite{Ni2008}.

\begin{figure}
\begin{center}
\includegraphics[width=8.8cm]{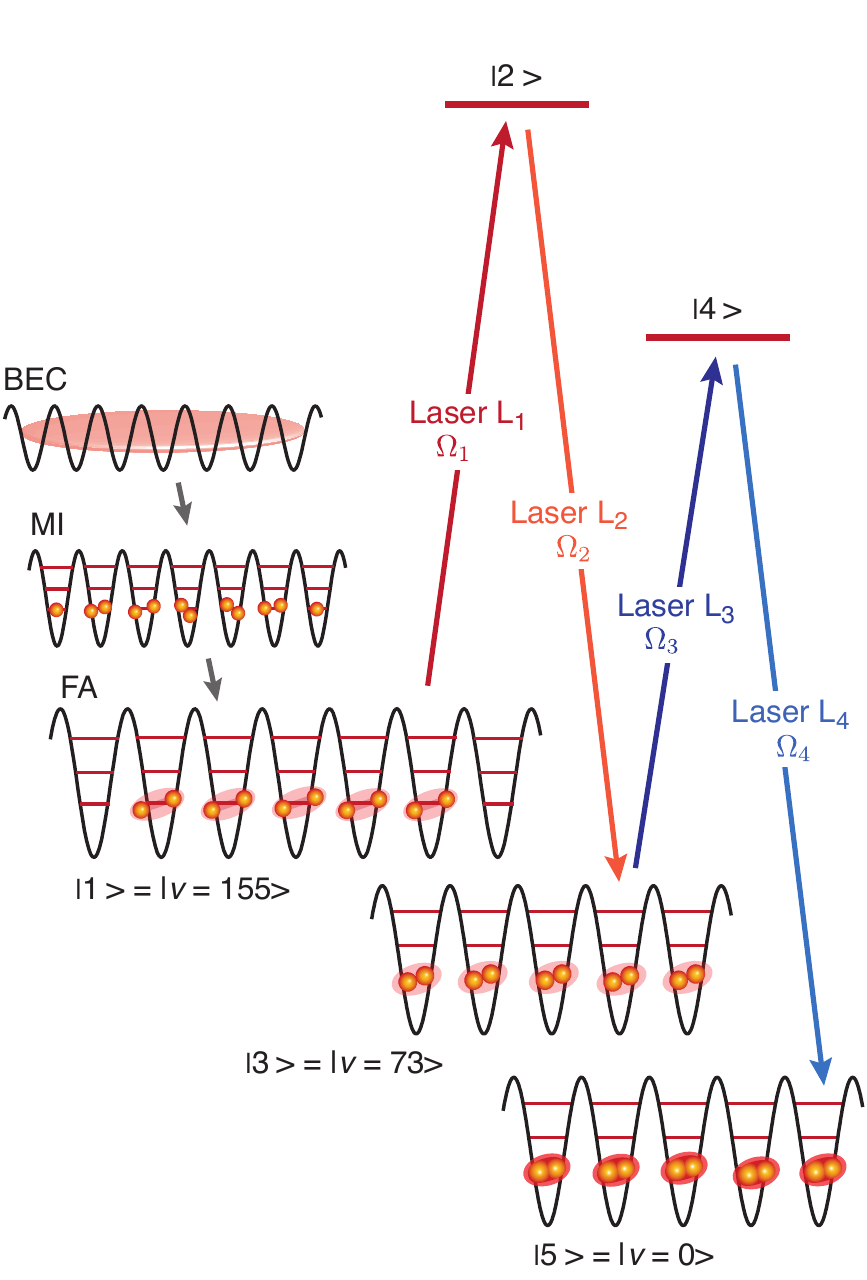}
\caption{}
 \label{fig1}
\end{center}
\end{figure}

Our molecular quantum-gas preparation procedure is summarized in Fig. \ref{fig1}. We load a BEC of Cs atoms\cite{Kraemer2004} into a three-dimensional optical lattice and drive the superfluid-to-Mott-insulator phase transition\cite{Bloch2008} under conditions that yield the highest number of lattice sites at which there are precisely two atoms (see the Methods section). We thus aim to maximize the size of the two-atom Mott shell\cite{Bloch2008}. With up to 45\% of the atoms at doubly-occupied lattice sites we come close to the theoretical limit of 53\% given the parabolic density profile of the BEC\cite{Volz2006}. The atom pairs reside in the motional ground state at each well and are then associated with near-unity probability\cite{Thalhammer2006} to Cs$_2$ Feshbach molecules, which are subsequently transferred to the weakly-bound level $|1\!>$, the starting level for the optical transfer (see the Methods section)\cite{Mark2007,Danzl2008Science,Danzl2009NJP}. Atoms at singly-occupied sites are removed by a combination of microwave and optical excitation\cite{Thalhammer2006}. We now have a pure molecular sample with near-unity occupation in the central region of the lattice. Each molecule is in the motional ground state of its respective well and perfectly shielded from collisional loss.

We employ stimulated Raman adiabatic passage (STIRAP)\cite{Bergmann1998} involving four laser transitions to coherently transfer the molecules into the lowest rovibrational level $|5\!> = |v\!=\!0,J\!=\!0\!>$ of the ground state singlet $X^1\Sigma_g^+$ potential as shown in Fig. \ref{fig2}a, bridging a binding energy of $h c \times$ 3628.7 cm$^{-1} \approx h \times 109$ THz\cite{Danzl2008Science}. Here, $v$ and $J$ are the vibrational and rotational quantum numbers, respectively, $h$ is Planck's constant and $c$ is the speed of light. For Cs$_2$, a homonuclear molecule, the four-photon process is preferred to a direct two-photon process because it allows us to overcome small Franck-Condon overlap. Lasers $L_1$ through $L_4$ couple $|1\!>$ and $|5\!>$ via three intermediate levels $|2\!>$,$|3\!>$, and $|4\!>$ (see the Methods section). For STIRAP in the presence of the lattice, the lattice light must not impede the transfer through optical excitation or by creating unwanted coherences. Also, the lattice wavelength has to be chosen such that the dynamical polarizabilities for $|1\!>$ and $|5\!>$ are closely matched in order to avoid excitation into higher motional states of the lattice as a result of motional wave-function mismatch\cite{Lang2008}. We typically set the lattice depth to a value of 20 $E_R$ for atoms, corresponding to $80 \; \tilde{E}_R$ for Feshbach molecules with twice the polarizability and double the mass and $83 \; \tilde{E}_R$ for molecules in $v=0$ at a lattice wavelength of 1064.5 nm, as determined below. Here, $E_R$ $(\tilde{E}_R)$ is the atomic (molecular) recoil energy.

Our experimental configuration ensures that only one particular molecular hyperfine sublevel is populated. The atomic BEC is prepared in the lowest hyperfine sublevel $|F_a\!=\!3,m_{Fa}\!=\!3\!>$, where $F_a$ and $m_{Fa}$ are the total atomic angular momentum and its projection on the magnetic field. Feshbach association and transfer between Feshbach levels via avoided crossings, as illustrated in  Fig. \ref{fig2}b (see the Methods section), conserve\cite{Chin2008} the total angular momentum projection $M_F=m_{Fa_1}+m_{Fa_2}=6$.
Fig. \ref{fig2}c shows the hyperfine structure of the target state, i.e. the rovibronic ground state $X^1\Sigma_g^+$ $|v\!=\!0,J\!=\!0\!>$. It splits into 28 hyperfine sublevels in the presence of a weak magnetic field, corresponding to the allowed values of the total nuclear spin $I$ = 0, 2, 4, 6 and its $2I+1$ projections $M_I$ for each value of $I$. The total energy splitting is $\sim h \times 270$ kHz at zero field\cite{Aldegunde2009} (see the Methods section). Importantly, there is only a single $M_I=M_F=6$ sublevel of $|v\!=\!0,J\!=\!0\!>$, namely the $|I=6,M_I=6>$ level. This level we selectively populate by exploiting the dipole selection rule $\Delta_{M_F}=0$ for linear polarization along the axis of quantization. It is the lowest-energy hyperfine sublevel and hence the absolute energy ground state of the Cs dimer for magnetic fields above $\sim13$ mT.

There are two possibilities for optical transfer from $ |1\!\!> $ to $ |5\!\!> $. Sequential STIRAP (s-STIRAP) uses two consecutive two-photon STIRAP processes, first from $ |1\!\!> $ to $ |3\!\!> $ and then from $ |3\!\!> $ to $ |5\!\!> $. The second scheme generalizes STIRAP\cite{Bergmann1998,Winkler2007} to the five-level system\cite{Malinovsky1997}: Four-photon STIRAP (4p-STIRAP) relies on the existence of a dark state of the form $|D\!\!> \ = (\Omega_2 \Omega_4 |1\!\!> - \ \Omega_1 \Omega_4 |3\!\!> + \ \Omega_1 \Omega_3 |5\!\!>)/A$ with time-dependent Rabi frequencies $\Omega_i\!=\!\Omega_i(t)$ for lasers $L_i$, $i\!=\!1,2,3,4$, and the appropriate normalization function $A\!=\!A(t)$. Similar to standard two-photon STIRAP, a counter-intuitive pulse sequence rotates the initial state $ |1\!\!> $ adiabatically into the final state, here $ |5\!\!> $. For this, $L_2$ and $L_3$ couple the three intermediate levels while $L_4$ and $L_1$ deliver time-dependent overlapping pulses with $L_4$ preceding $L_1$. Fig. \ref{fig3}b and e show the timings for both schemes including the reverse sequence used for detecting the ground-state molecules after a certain hold time $\tau_h$.

We investigate 4p-STIRAP to $|v\!=\!0,J\!=\!0\!>$ by interrupting the transfer sequence after a given 4p-STIRAP time $\tau$ and measuring the number of Feshbach molecules, as shown in Fig. \ref{fig3}a. The molecules are transferred to $|5\!>$ in a single step. No molecules in $|1\!>$ are detected during $\tau_h$ as the remaining Feshbach molecules are cleared by $L_1$ at the end of the transfer. When the pulse sequence is reversed, a large fraction of the molecules returns to $|1\!>$. Typically, 30\% of the molecules are recovered after the full double 4p-STIRAP sequence. Assuming equal efficiencies for both transfers, the single-pass efficiency is 55\%. The inset shows the double 4p-STIRAP efficiency versus detuning $\Delta_4$ of $L_4$ from the ($|4\!> \rightarrow |5\!>$)-transition with all other lasers on resonance. With ground-state transfer efficiencies between 55\% and 60\%, about half of the lattice sites are occupied by a ground-state molecule. The solid lines in Fig. \ref{fig3}a represent a simulation of 4p-STIRAP that takes into account excited-state spontaneous decay and laser linewidth. Transfer times are typically 4 $\mu$s to 10 $\mu$s. The simulation yields that the transfer efficiency is currently limited by a combination of laser linewidth, which is about 10 kHz when averaged over 1 s, and imperfect adiabaticity due to finite available laser power to drive the extremely weak transitions of the 5-level scheme\cite{Danzl2008Science,Mark2009}. Molecules not transferred to $|5\!>$ as a result of insufficient phase coherence or limited adiabaticity are excited to either $|2\!>$ or $|4\!>$ by one of the lasers and are hence pumped into a multitude of rovibrational levels, which do not couple to the rovibrational ground state.
For comparison, the double s-STIRAP efficiency from $|3\!>$ to $|v\!=\!0,J\!=\!0\!>$ and $|v\!=\!0,J\!=\!2\!>$ is shown in Fig. \ref{fig3}c and d, respectively. The solid lines represent a calculation matched to the data for standard 3-level STIRAP. With 55\%-60\%, the total ($|1\!>\rightarrow |5\!>$)-transfer efficiency for s-STIRAP is comparable to 4p-STIRAP.

A crucial prerequisite for efficient ground-state transfer without heating is perfect matching of the motional wave functions for the initial weakly-bound state and the final ground state. A mismatch leads to unwanted excitation of higher lattice vibrational levels or bands and hence to loss of state control. The lattice thus has to be operated at the magic wavelength condition\cite{Ye2008}, i.e. at a wavelength that gives equal light shifts for the initial and the final molecular states. We measure the lattice band structure and determine the molecular polarizability of the ground-state molecules (see the Methods section). Molecules residing in the lowest band of the lattice are excited to the first (second) band by phase (amplitude) modulation of the light generating the lattice. Fig. \ref{fig4} shows the measured band energies together with a calculation of the band structure as a function of lattice depth. On resonance, excitation to higher bands can readily be observed in momentum space as shown in inset {\textbf a}. For comparison, off-resonant modulation transfers hardly any population into higher bands (see inset {\textbf b}). We determine the band energies by taking modulation spectra as shown in inset {\textbf c}. We then use the band structure calculation to fit all measured resonance positions with the molecular dynamical polarizability $P_{|v=0>}$ as the single free parameter. These measurements are done for $|v\!=\!0,J\!=\!2\!>$. We obtain $P_{|v=0>}= 2.1(1) \times P_a$, where $P_a$ is the atomic polarizability. For the initial, weakly-bound Feshbach molecules in level $|g\!>$ we obtain $P_{|g>}= 2.0(1) \times P_a$. Hence the magic wavelength condition is well fulfilled.

We measure the lifetime $\tau$ of the molecules in the lattice by varying the hold time $\tau_h$ for up to $20$ s and recording the number of remaining molecules as shown in Fig. \ref{fig5}. To reduce inelastic light scattering, the lattice depth was adiabatically reduced to about 41.5 $\tilde{E}_R$ after the 4p-STIRAP transfer. An exponential fit gives a lifetime of $\tau=8.1(6)$ s. We attribute this long lifetime to the large detuning $\Delta_L \approx 6.9$ THz from the lowest $0_u^+$ level with predominant $A^1\Sigma_g^+$ singlet contribution as shown in the inset to Fig. \ref{fig5}. Levels of the $0_u^+$ system that lie below this are almost purely of $b^3\Pi_u$ character and thus make negligible contributions to the optical excitation rate.

We are now in a position to determine collisional properties of ultracold ground-state molecules in a fully state-selective way. At magnetic fields beyond 13 mT, where the level $|I=6,M_I=6>$ becomes the absolute ground state, the sample should show collisional stability and thus allow the formation of a BEC of ground-state molecules when the lattice is adiabatically removed\cite{Jaksch2002}. The long coherence times and the perfect decoupling from the environment in an optical lattice as demonstrated here will enable a new generation of precision measurements\cite{DeMille2008est,Zelevinsky2008pto}. Furthermore, our results can readily be generalized to heteronuclear systems such as KRb\cite{Ni2008} and RbCs\cite{Pilch2009}, opening up the possibility to study dipolar quantum phases in optical lattices.

\section{Methods}

\subsection{Lattice loading}
We first follow the procedure detailed in Ref.\cite{Danzl2009NJP}. In brief, we produce an atomic BEC with typically $1\times10^5 $ Cs atoms in the lowest hyperfine sublevel $F_a\!=\!3, \ m_{Fa}\!=\!3$ in a crossed optical dipole trap. We then drive the superfluid-to-Mott-insulator phase transition\cite{Bloch2008} by exponentially ramping up the power in a three-dimensional optical lattice
within about $300$ ms. The lattice is generated by three mutually-orthogonal,
retro-reflected laser beams at a wavelength of $ \lambda= 1064.5$ nm, each with a $1/e$-waist of about $350 \ \mu$m. While ramping up the lattice potential, the power in the two dipole-trap beams is increased to ensure that the central density in the trap is sufficiently high to allow formation of atom pairs at the central wells of the lattice, but not too high to lead to triply occupied sites. We ramp the lattice to a depth of about 20 $E_R$ before Feshbach association. Here, $E_R=h^2/(2 m_a \lambda^2) = k_B \! \times \! 64 \, $nK is the atomic photon-recoil energy with the mass $m_a$ of the Cs atom and Boltzmann's constant $k_B$. Up to 45\% of the atoms reside at doubly occupied lattice sites. We estimate this number from the molecule production efficiency. This value is close to the maximum of 53\% that can be achieved for loading a harmonically confined BEC in the Thomas-Fermi limit into an optical lattice\cite{Volz2006}.

For the molecules, the recoil energy is $\tilde{E}_R=h^2/(2\times 2m_a \lambda^2)$. The polarizability of Feshbach molecules is twice the atomic polarizability. The same lattice that has a depth of 20 $E_R$ for the atoms has thus a depth of 80 $\tilde{E}_R$ for the Feshbach molecules.

\subsection{Feshbach association and Feshbach state transfer}

We efficiently produce weakly bound Cs$_2$ Feshbach molecules in the presence of the optical lattice by a magnetic field sweep\cite{Chin2008} across a narrow g-wave Feshbach resonance with its pole at a magnetic field value of $B=1.98$ mT\cite{Herbig2003,Mark2007}. The molecules are initially in level $|g\!>$, for which $\ell \! = \! 4$. Here, $\ell$ is the quantum number associated with the mechanical rotation of the nuclei\cite{Chin2008}. We subsequently transfer the molecules via level $|g_2\!>$ with 95\% efficiency into level $|s\!> \equiv |1\!> $ with $\ell \! = \! 0$ by magnetic field ramping\cite{Mark2007,Danzl2009NJP} as shown in Fig. \ref{fig2}b. For this level, the transitions to excited molecular levels are stronger than for the initial level $|g\!>$\cite{Danzl2009Faraday}. We obtain up to 2.5 $\times 10^4$ Feshbach molecules in the lattice in the desired starting state. We detect the molecules in $ |1\!> $ by reversing the Feshbach state transfer sequence, dissociating the molecules at the Feshbach resonance and detecting the resulting atoms by standard absorption imaging\cite{Herbig2003}.

\subsection{Molecular states for ground state transfer}

The relevant molecular states for Cs$_2$ are shown in Fig. \ref{fig2}a. Levels
$|2\!>$ and $|4\!>$ belong to the coupled $(A^1\Sigma_g^+-b^3\Pi_u)0_u^+$ potentials\cite{Danzl2008Science}. We have recently identified suitable transitions linking $|1\!>$ to $|5\!>$, where levels $|2 >$, $|3 >$, and $|4 >$ were chosen to give balanced transition strengths on the four optical transitions\cite{Danzl2009Faraday,Mark2009}. For $|3\!>$ we choose either $|v\!=\!73,J\!=\!2\!>$ or $|v\!=\!73,J\!=\!0\!>$ of the $X^1\Sigma_g^+$ ground state with a binding energy of $\sim h c \times 1061$ cm$^{-1}$.

\subsection{Hyperfine structure of the rovibronic ground state}

The hyperfine levels are calculated using the molecular constants from Ref. \cite{Aldegunde2009} by constructing and diagonalizing a Hamiltonian matrix in an uncoupled basis set of functions representing the molecular rotation and the spins of the two nuclei, using the matrix elements given in the Appendix of Ref. \cite{Aldegunde2009}. For $J=0$ states the hyperfine structure is dominated by the scalar spin-spin coupling and the nuclear Zeeman effect, but for $J>0$ additional terms are important.

\subsection{STIRAP laser setup}

STIRAP lasers $L_i$ with $i=1,2,3,4$ are continuous-wave grating-stabilized tunable diode lasers, which are stabilized to optical resonators for short-term stability and referenced to an optical frequency comb for long-term stability and reproducibility. We estimate the linewidth of the lasers to be about 10 kHz. In order to ensure minimum momentum recoil imparted on the molecules, the beams for lasers $L_1$ and $L_2$ are co-propagating. The beams for $L_3$ and $L_4$ are also co-propagating but run antiparallel to the beams of $L_1$ and $L_2$. All beams run horizontally and are linearly polarized with the polarization axis in the vertical direction, parallel to the direction of the magnetic field, which defines the axis of quantization. We operate at Rabi frequencies in the range of $2\pi \times$ (1 to 4) MHz.

\subsection{Polarizability measurement}

For determining the ground state molecular polarizability, transfer to $|v=0>$ is performed at a fixed lattice depth of 83 $\tilde{E}_R$ for $|v=0>$ molecules. The lattice depth is then ramped to the desired value within 50 ms. For phase modulation of the lattice, the frequency of one lattice beam is usually modulated with a modulation depth of 2 MHz at the desired frequency for about 10 ms. For amplitude modulation, the intensity is typically modulated by 20\% for about 10 ms.

\begin{addendum}

\item[Acknowledgements] We thank H. Ritsch, S. D\"urr, N. Bouloufa, and O. Dulieu for valuable discussions. We are indebted to R. Grimm for generous support and to H. H\"affner for the loan of a charge-coupled camera. We gratefully acknowledge funding by the Austrian Ministry of Science and Research (Bundesministerium f\"ur Wissenschaft und Forschung) and the Austrian Science Fund (Fonds zur F\"orderung der wissenschaftlichen Forschung) in form of a START prize grant and by the European Science Foundation within the framework of the EuroQUASAR collective research project QuDeGPM and within the framework of the EuroQUAM collective research project QuDipMol. R.H. is supported by a Marie Curie International Incoming Fellowship within the 7th European Community Framework Programme.
 
\item[Competing Interests] The authors declare that they have no
competing financial interests.

\item[Correspondence] Correspondence and requests for materials
should be addressed to H.-C. N. (email: christoph.naegerl@uibk.ac.at) or to J. G. D. (email: johann.danzl@uibk.ac.at).

\end{addendum}

\newpage

\newpage
\clearpage
Figure 1:
{\bf Molecular quantum gas preparation procedure.} A BEC of Cs atoms is loaded into an optical lattice. By increasing the lattice depth, a Mott-insulator state (MI) with preferentially two atoms per site is created. Feshbach association (FA) subsequently converts atom pairs into weakly-bound molecules in state $|1\!\!>$. These are then transferred in the presence of the lattice to a specific hyperfine level $|I\!=\!6, M_I\!=\!6\!\!>$ of the rovibronic ground state $|5\!\!>$ = $X^1\Sigma_g^+$ $|v\!=\!0,J\!=\!0\!\!>$ by a stimulated four-photon process (STIRAP) involving lasers $L_i$ with Rabi frequencies $\Omega_i$ $i=1,2,3,4$, and three intermediate levels $|2\!\!>$, $|3\!\!>$, and $|4\!\!>$.

\newpage

\begin{figure}
\begin{center}
\includegraphics{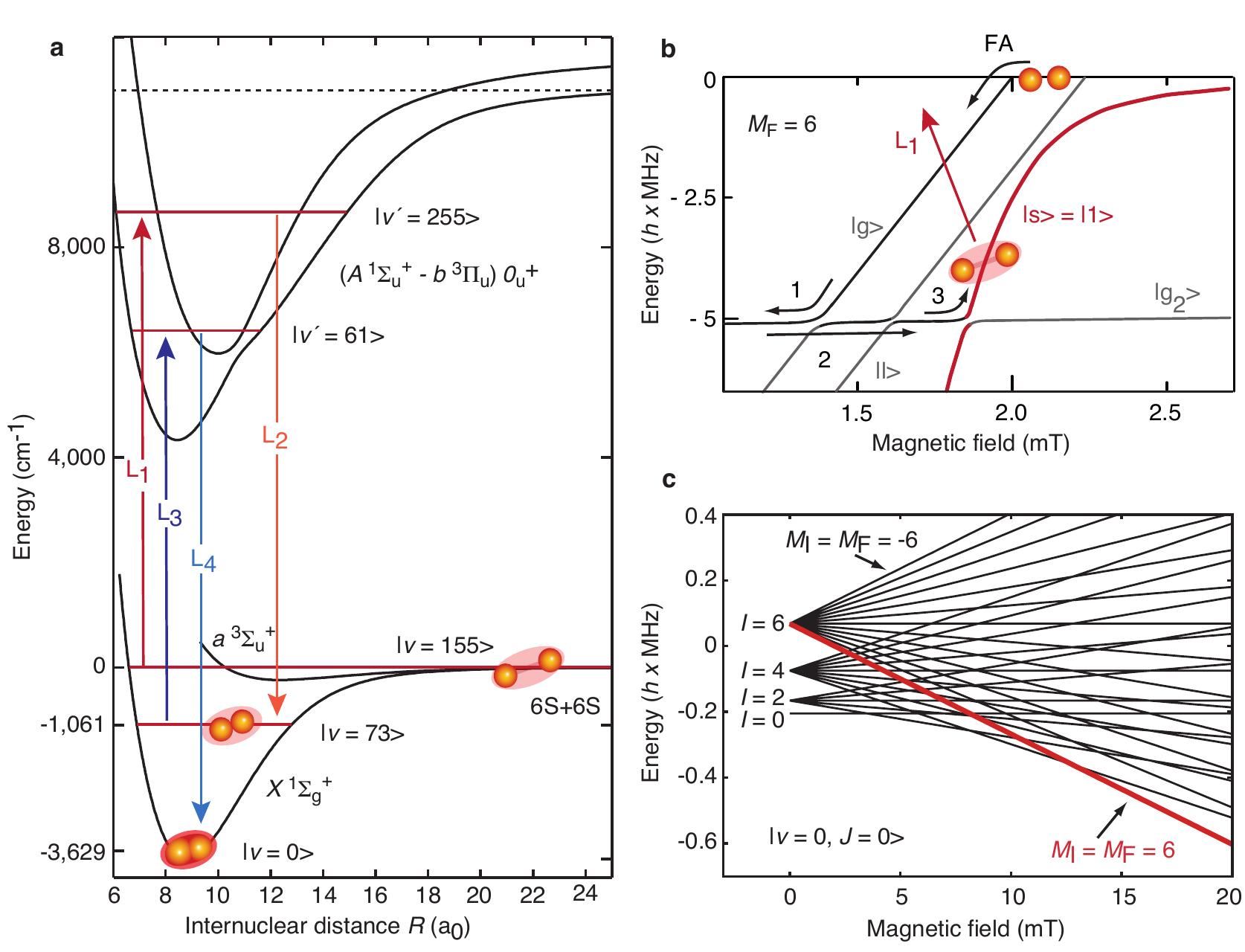}
\caption{ }
\label{fig2}
\end{center}
\end{figure}

\newpage
\clearpage

Figure 2:
{\bf Molecular potentials and level schemes for ground-state transfer.} {\textbf a,} The four-photon transfer from the weakly-bound Feshbach level $|1\!\!>= |\nu\!\approx\!155>$ (not resolved near the 6S+6S asymptote) to the rovibrational ground state $|5\!\!>=|\nu\!=\!0,J\!=\!0\!\!>$ of the singlet $X^1\Sigma_g^+$ potential involves the deeply bound level $|3\!\!> = |\nu\!=\!73>$ of the $X^1\Sigma_g^+$ potential\cite{Danzl2008Science} and the levels $|2\!\!>=|\nu'\!=\!225,J\!=\!1\!\!>$ and $|4\!\!>=|\nu'\!=\!61,J\!=\!1\!\!>$  of the electronically excited $(A^1\Sigma_u^+ - b^3\Pi_u) \ 0_u^+$ potentials\cite{Danzl2009Faraday,Mark2009}. The laser wavelengths for $L_1$, $L_2$, $L_3$, and $L_4$ are near 1126 nm, 1006 nm, 1351 nm, and 1003 nm, respectively. {\textbf b,} Zeeman diagram for weakly bound molecules near the 6S+6S asymptote. Molecules are associated at a $g$-wave Feshbach resonance\cite{Herbig2003} at 1.98 mT (FA) and then transferred to the desired starting level $|1\!\!>=|s\!\!>$ for optical transfer via three avoided level crossings by slow (arrows 1,3) and fast (arrow 2) magnetic field ramps\cite{Mark2007}. The binding energy is given with respect to the $(F_{a_1}\!=\!3, m_{Fa_1}\!=\!3)\times(F_{a_2}\!=\!3, m_{Fa_2}\!=\!3)$ two-atom lowest hyperfine asymptote. All Feshbach levels are characterized by $M_F=6$. {\textbf c,} Calculated Zeeman diagram for the hyperfine manifold of the rovibronic ground state $|5\!\!> = |\nu\!=\!0,J\!=\!0\!\!>$. The optical transfer goes selectively to level $|I\!=\!6, M_I\!=\!6\!\!>$, indicated in red. This level becomes the lowest-energy absolute ground state for magnetic-field values above $\sim 13$ mT. There are no avoided crossings between different hyperfine sublevels\cite{Aldegunde2009}.

\newpage
\begin{figure}
\begin{center}
\includegraphics[width=17cm]{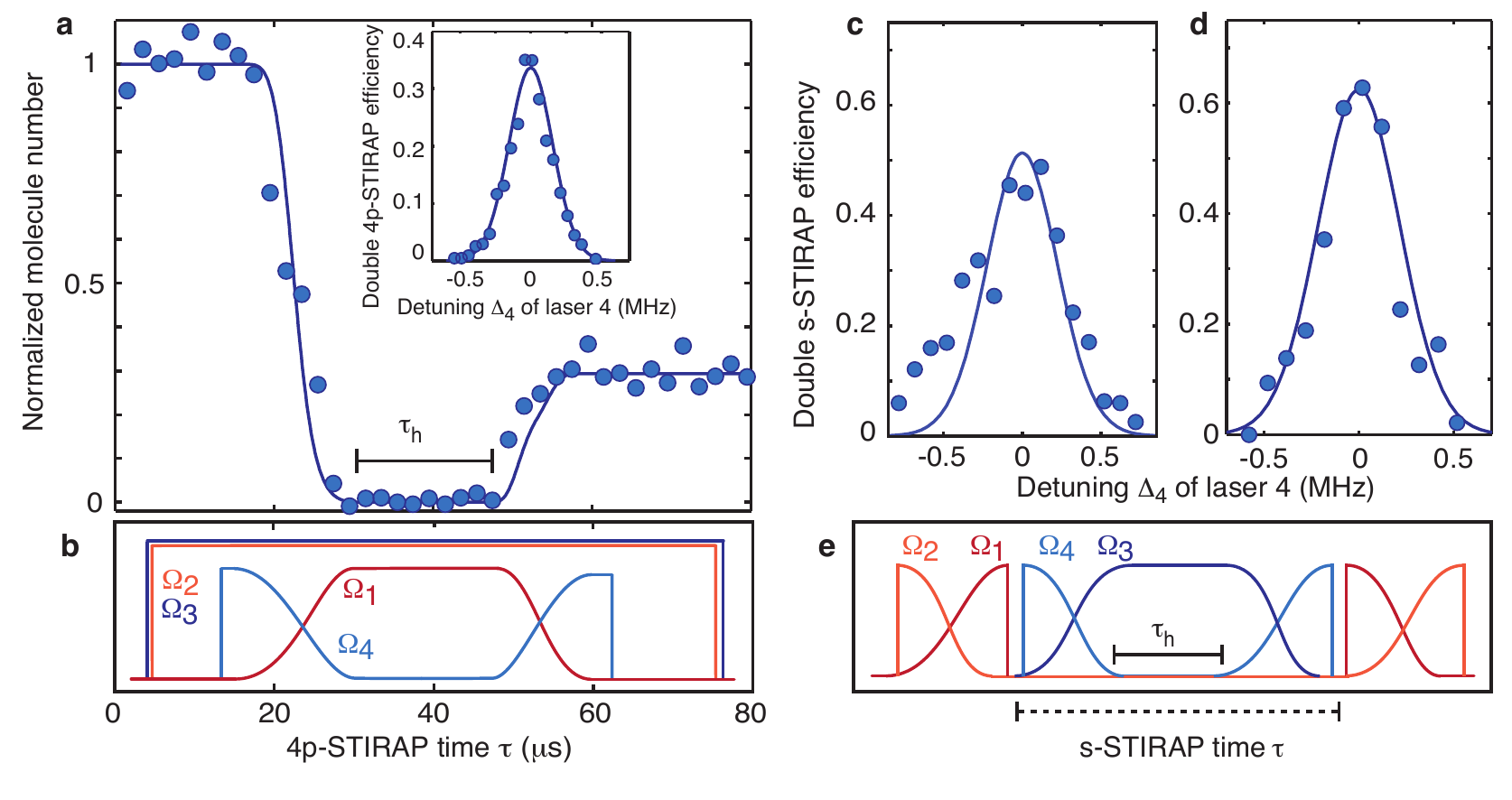}
\caption{}
\label{fig3}
\end{center}
\end{figure}

\newpage
\clearpage
Figure 3:
{\bf STIRAP transfer to the rovibronic ground state $|5\!\!>=|\nu\!=\!0, J\!=\!0\!\!>$ and back.} {\textbf a,} 4p-STIRAP transfer and {\textbf b,} schematic timing for the Rabi frequencies $\Omega_i$, $i=1,2,3,4$: Number of molecules in state $|1\!\!>$ as a function of 4p-STIRAP time $\tau$ for all 4 lasers on resonance. The lattice depth is 80 $\tilde{E}_R$ and 83 $\tilde{E}_R$ for molecules in levels $|1\!\!>$ and $|5\!\!>$, respectively. Data points represent a single experimental realization, not an average over several runs. The solid line is a 4p-STIRAP model calculation. $\tau_h$ is the hold time in $|5\!\!>=|\nu\!=\!0, J\!=\!0\!\!>$.
Inset: double 4p-STIRAP efficiency as a function of the detuning $\Delta_4$ of laser $L_4$ and corresponding model calculation. The peak corresponds to a single-pass efficiency of 57\%. {\textbf c,} and {\textbf d,} s-STIRAP: Double STIRAP efficiency for the inner two-photon STIRAP from $|3\!>$ to $|v\!=\!0,J\!=\!0\!>$ ({\textbf c}) and to $|v\!=\!0,J\!=\!2\!>$ ({\textbf d}) and back, corresponding to the dotted bar in the timing sequence in {\textbf e}, as a function of the detuning $\Delta_4$ of laser $L_4$. The number of molecules is normalized to the initial number in $|3\!>$. All measurements are performed at an offset magnetic field of $1.9$ mT.

\newpage
\begin{figure}
\begin{center}
\includegraphics[width=8.8cm]{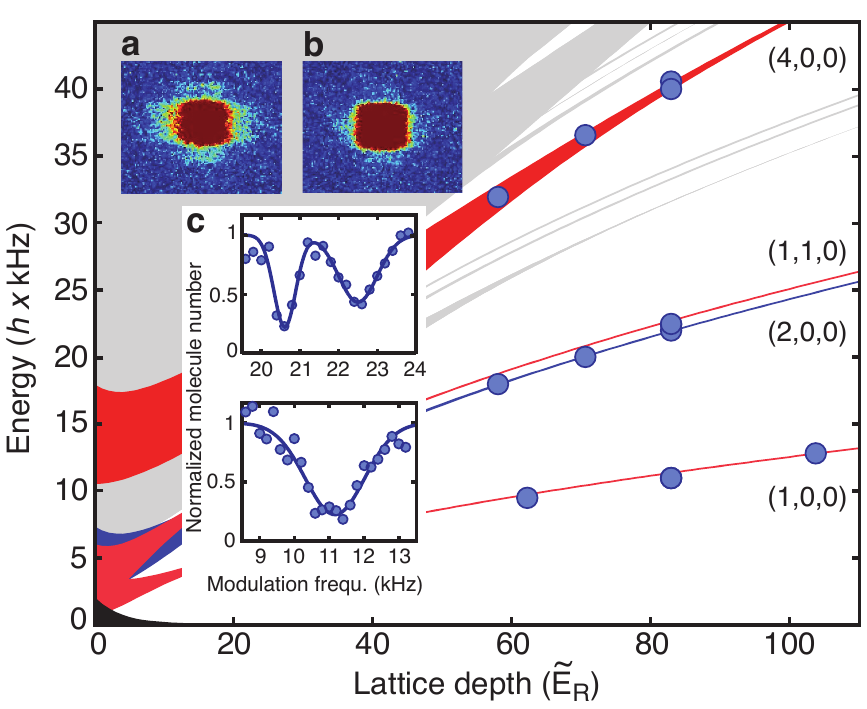}
\caption{}
\label{fig4}
\end{center}
\end{figure}

\newpage
\clearpage
Figure 4:
{\bf Lattice band structure for $|v\!=\!0\!>$ molecules.} Band energies as a function of lattice depth in units of the molecular recoil energy $\tilde{E}_R$ as measured by phase and amplitude modulation of the lattice. The lattice bands are labeled by $(k,l,m)$, where $k,l,$ and $m$ give the number of vibrational quanta along the three spatial directions in the limit of a deep lattice. The horizontal position of the data points (filled circles, representing the position of excitation resonances as shown in inset {\bf c}) is given by the molecular polarizability, which is determined by a fit of the data  to the band structure. Inset {\bf a} shows the molecular momentum distribution after transfer to higher lattice bands by resonant lattice amplitude modulation. The distribution represents an average of 5 experimental runs, smoothed with a Gaussian filter. For comparison, inset {\bf b} shows that hardly any population is transferred to higher bands for off-resonant modulation. The rectangular shape of the first Brillouin zone is easily seen. Inset {\bf c} shows typical excitation spectra for amplitude (top) and phase (bottom) modulation at 83 $\tilde{E}_R $. For these, we determine the number of molecules in the first Brillouin zone as a function of the excitation frequency. The solid lines are Gaussian fits. The resonance at 22.5 kHz corresponds to excitation to the nearly-degenerate bands (2,0,0) and (1,1,0) (not resolved). The resonance at 20.6 kHz is a two-phonon excitation to (4,0,0).

\newpage
\begin{figure}
\begin{center}
\includegraphics{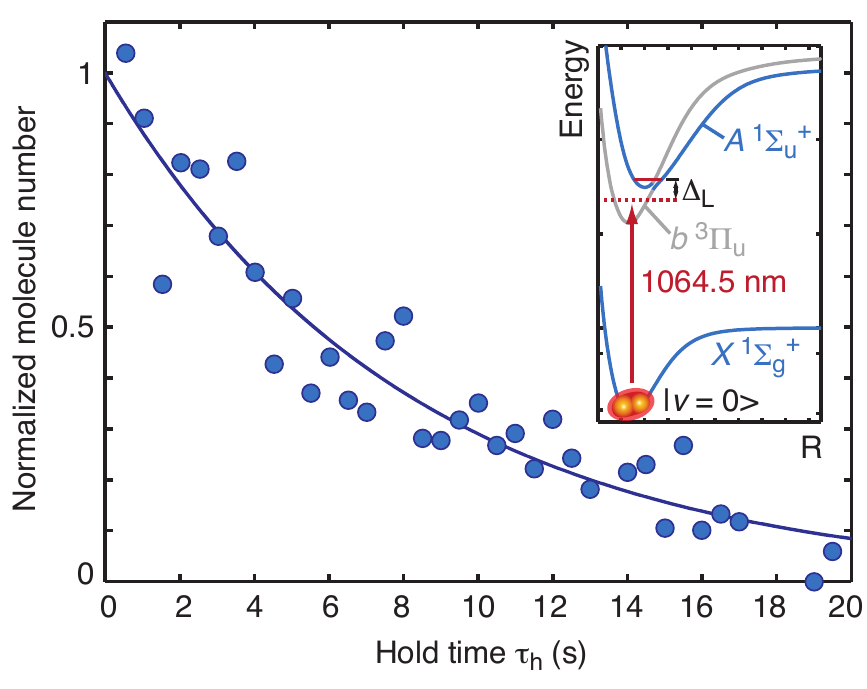}
\caption{} 
\label{fig5}
\end{center}
\end{figure}

\newpage
\clearpage
Figure 5:
{\bf Lifetime of trapped ground-state molecules in the optical lattice.} Normalized number of molecules in $|5\!\!>=|\nu\!=\!0, J\!=\!0\!\!>$ as a function of hold time $\tau_h$. The solid line is an exponential fit, yielding a lifetime of $8.1(6)$s. The inset schematically shows the excited-state potentials to which off-resonant optical excitation is possible (cf. Fig. \ref{fig2}a). $\Delta_L$ is the detuning of the lattice light at 1064.5 nm with respect to the lowest $0_u^+$ level with $A^1\Sigma_u^+$ character. During the hold time, all STIRAP laser fields are turned off.

\end{document}